\documentclass[%
  prb,% 
  letterpaper,%
  twocolumn,%
  showpacs,%
  floatfix,%
  superscriptaddress%
]{revtex4-1}

\RequirePackage{graphicx,amsmath,amssymb,bm}
\usepackage{lineno}
\RequirePackage{hyperref}
\hypersetup{%
  breaklinks = {true},
  citecolor = {blue},
  colorlinks = {true},
  linkcolor = {red},
  pdfauthor = {\textcopyright\ Carlos Paez},
  pdfcreator = {\LaTeX\ and \flqq hyperref\frqq},
  pdffitwindow = {true},
  pdfmenubar = {true},
  pdfpagelayout = {SinglePage},
  pdfstartview = {Fit},
  pdftoolbar = {true},
  plainpages = {false},
}

\newcommand{\Fref}[1]{Fig.~\ref{#1}}

\renewcommand{\eqref}[1]{eq.~(\ref{#1})}

\RequirePackage[usenames,dvipsnames]{xcolor}
\usepackage{soul}

\newcommand{\unicamplimeira}{Faculdade de Ci\^{e}ncias Aplicadas, Universidade
Estadual de Campinas, 13484-350 Limeira, SP Brazil}

\newcommand{\mackgrafe}{ MackGraphe -Graphene and Nano-Materials Research Center, Mackenzie
Presbyterian University,
Rua da Consola\c{c}\~{a}o 896, 01302-907, S\~{a}o Paulo, SP, Brazil}

\begin{document}

%\preprint{APS/123-QED}
%\linenumbers
\title{Current flow in biased bilayer graphene: the role of sublattices}
% Force line breaks with \\
%\thanks{A footnote to the article title}%

\author{C. J. P\'aez}
\email[Corresponding author:
]{carlos.gonzalez@fca.unicamp.br}
\affiliation{\unicamplimeira}%
\author{D. A.  Bahamon}%
\affiliation{\mackgrafe}%
\author{Ana L. C. Pereira}%
\affiliation{\unicamplimeira}%

\date{\today}% It is always \today, today,
             %  but any date may be explicitly specified

\begin{abstract}
We investigate here how the current flows over a bilayer graphene in the presence of an external
electric field perpendicularly applied (biased bilayer).  Charge density polarization between 
layers in these systems is known to create a layer pseudospin, which can be manipulated by 
the electric field. Our results show that current does not necessarily flow over regions of the 
system with higher charge density. Charge can be
predominantly concentrated over one layer, while current flows over the other layer. 
We find that this phenomenon occurs when the charge density becomes highly concentrated over 
only one of the sublattices, as the electric field breaks layer and sublattice symmetries 
for a Bernal-stacked bilayer. For bilayer nanoribbons, the situation is even more complex, with 
a competition between edge and bulk effects for the definition of the current flow. 
We show that, in spite of not flowing trough the layer where charge is polarized to, the current
in these systems also defines a controllable layer pseudospin. 

\end{abstract}

% PACS, the Physics and Astronomy
                             % Classification Scheme.
%\keywords{Suggested keywords}%Use showkeys class option if keyword
                              %display desired
\maketitle

%\tableofcontents

%================================================================
\section{Introduction}
\label{sec:Intro}
%================================================================

For electronic devices of reduced dimensions, the spatial mapping of charge current is of paramount
importance. In a quantum point contact, for example, electrons flow through narrow branching
channels rather than the expected uniform propagation\cite{TopLS00,TopLW01}; these measurements
are crucial to understand how the geometry and impurities of the device affect its performance.
Graphene, due to its exceptional electronic properties, has been pointed out to have great
potential to replace existing materials in traditional electronics \cite{CasGPN09}, 
as well as to be used in new
pseudospintronic devices\cite{RycTB07,SanPM09,AkhB07,GunW11,Sch10,PesM12}. Common to these
traditional and new applications, local aspects of charge flow in graphene have to be understood.
Studies on zigzag graphene nanoribbons have shown, for low energies, dispersionless and sublattice
polarized edge states\cite{CasPL08}. Notably the overlap of these no current-carrying states from
opposite edges creates charge flow through the centre of the nanoribbon\cite{ZarN07}. This
charge-current asymmetry has been ignored for bilayer graphene (BLG), which offers better options
for digital electronics.

A remarkable property of BLG -which potentiates its use in future graphene based electronics- is
the possibility of opening and controlling a band gap with a
potential difference applied between top and bottom layers (biased bilayer)
\cite{McC06,CasNN10,Sch10,DonVJ13,OhtBA06,NilCG08,McCK13}. The
externally applied perpendicular electric field breaks the inversion symmetry of the system
\cite{ZhaTG09} allowing to define a layer pseudospin,  at least for energies below the interlayer
coupling energy \cite{PesM12,AbdPM07,SanPM09}. Therefore, many devices based on these systems have
been proposed recently, which involve the ability to control this layer pseudospin (the charge
density polarization between layers induced by the bias) for different bias
layouts\cite{MiyTK10,XiaFL10,Vel12}, such as the creation of electron highways \cite{QiaJN11} or
pseudospin-valve devices \cite{RycTB07,SanPM09,LiZX10}.
Experimentally, charge localization over different layers and different sublattices due to a bias
voltage has been observed in these systems by STM images\cite{KimKW13}, indicating the possibility
of controlling layer and sublattice pseudospins in real samples.
Even though all the attention that has been given to the possibilities of controlling charge
densities in BLG through the bias, the charge flow has been neglected.

The purpose of this paper is to analyze and quantify the main transport features of pristine biased
BLG. In particular we are enticed to unveil the relation between charge density and charge flow.
Although charge and current are intimately linked by the continuity equation, when the electric
field localizes charge over different sublattices in different layers, it is not evident how
current
density is distributed. Using the lattice Green's functions we are able to map charge density over
each sublattice site in both layers, conjointly with the current flowing towards its neighbors.
Our results show that current does not necessarily flow in the regions with higher charge density,
and this would have a fundamental role when devising electronics.  At low energies, for bulk biased
BLG, we observe that charge is primarily concentrated over one layer while current flows over the
other layer. We show that this is a consequence of an important concentration of charge in only one
of the sublattices in the layer  with more charge density.
This picture is enriched in biased BLG nanoribbon with zigzag edges
where additional sublattice polarized edge states \cite{NakFD96,LimSF10,LvL12,WanSL11} 
compete with bulk sublattice polarization. The distribution of the current for edge states 
is found to also depend whether the most external atom of the edge
corresponds to a coupled or uncoupled sublattice in the AB stacking. We show results as a function
of energy around Fermi energy, and also as a function of nanoribbon width and bias strength ($V$),
elucidating the behavior of current flow and the main role of sublattices. The effects of disorder 
and next-nearest neighbor hoppings are also discussed.

\begin{figure}[h]
\begin{center}
%\vskip2.0cm
\includegraphics[width=0.95\columnwidth]{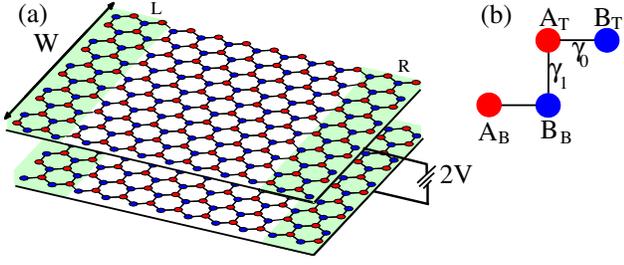}
\caption{(color online)(a) Schematic representation of a BLG nanoribbon, with
zigzag edges and width W, between left (L) and right (R) semi-infinite contacts.
There is an electrostatic potential difference of 2V between the two layers.
Different sublattices, A and B, are represented in different colors (b)
Detail of the sublattices A and B in top and bottom layer,
indicating of the nearest hoppings:
$\gamma_0$ in-plane, and $\gamma_1$ coupling the dimer sites interlayer.}
\label{fig:device}
\end{center}
\end{figure}

%================================================================
\section{Model}
\label{sec:theory}
%================================================================

We consider a BLG nanorribon with zigzag edges and Bernal AB stacking, as shown in 
\Fref{fig:device}(a). The width
$W=(N-1)a\frac{\sqrt{3}}{2}$ is defined by the number of sites $N$ in the transversal direction and
$a=2.46$ {\AA} is the lattice constant for graphene. The infinite BLG zigzag nanoribbon is modelled 
by
the tight binding Hamiltonian:

\begin{equation}
\label{eq:hamiltonian}
\begin{array}{c}
H=-\gamma_0\displaystyle\sum_{m,i,j}(a_{m,i}^{\dagger}b_{m,j}+H.c.)
-\gamma_1\displaystyle\sum_{j}(a_{T,j}^
{\dagger}
b_{B,j}+H.c.)\\
+ V \displaystyle\sum_{i}(a_{T,i}^{\dagger}a_{T,i}+b_{T,i}^{\dagger}b_{T,i})-V
\displaystyle\sum_{i}(a_{B,i}^{\dagger}a_{B,i}+b_{B,i}^{\dagger}b_{B,i}),
\end{array}
\end{equation}

Here,  the first term refers to  individual graphene layer (top and bottom), the second term
describes the interlayer coupling, and  the last two terms introduce the interlayer bias which 
induces 
an energy difference between layers parameterized by $V$. Field
operators $a^{\dagger}_{m,i}$ ($a_{m,i}$), $b^{\dagger}_{m,i}$ ($b_{m,i}$) create (annihilate)  one
electron in sublattice A or B \emph{i-th} site of the top ($m=T$) or bottom ($m=B$) layer. We use
the  intralayer nearest-neighbor hopping $\gamma_0=3.16$ eV and the interlayer coupling $\gamma_1$=
0.381 eV \cite{KuzCV09}. From our Hamiltonian \eqref{eq:hamiltonian} the Bernal AB stacking is
easily recognised, as shown in \Fref{fig:device}(b), sublattice sites A in the top layer ($A_T$)
are on top of sublattice sites B in the bottom layer ($B_B$). We refer to this 
coupled($A_T-B_B$) sites as dimer sites while non-coupled sites ($A_B$ or $B_T$) are non-dimer 
sites \cite{CasPL08,McCK13}. The Introduction of next-nearest neighbor hopping in each layer and 
further interlayer couplings, as well as of on site disorder, is discussed in section VI.

To account for the electronic transport properties, the infinite BLG zigzag nanoribbon is divided in
three regions; a finite central region and two semi-infinite ribbons acting as contacts
\cite{Datta99}. Although \eqref{eq:hamiltonian} describes the dynamics of electrons in the three
regions and no qualitative differences can  be found among them; it is mandatory to have a finite
central region to calculate its Green's function \cite{Datta99,FG99,LewM13} in order to extract
conductance, local density of states (LDOS) ($\rho$), charge density ($\rho_c$) and current density
($\vec{J}$)\cite{BahPS11}. Despite the fact the transport properties are calculated for the central 
region, these
can be extended to the whole BLG zigzag nanoribbon. The retarded Green's function is calculated as
$G^r=[E-H_c-\Sigma_L-\Sigma_R]^{-1}$ where $H_c$ is the Hamiltonian of the central region and
$\Sigma_{L(R)}$ is the left (right) contact self-energy \cite{Datta99}.

Charge and current are intimately related through the continuity equation,  its lattice version  can
be written as $\frac{dc^{\dagger}_nc_n}{dt} + [\hat{J}_{nn'}-\hat{J}_{n'n}]=0$ where
$\hat{J}_{nn'}=\frac{e}{i\hbar}[t_{n'n}c^{\dagger}_{n'}c_n-t_{nn'}c^{\dagger}_nc_{n'}]$ is the bond
charge current operator. $\hat{J}_{nn'}$ results, exactly as one would expect, from the difference
of electrons flow in opposite directions. The connection with the Green's function arises because
the quantum statistical average of the bond charge current operator of the form $\langle
c^{\dagger}_{n}c_{n'} \rangle$ are related to the lesser Green's function $G^<_{n'n}(E)$
\cite{HauJ08,Datta99}, in steady state the bond charge current including spin degeneracy is:

\begin{equation}
  J_{nn'}=I_0\int_{E_F-eV/2}^{E_F + eV/2}
  dE[t_{n'n}G_{nn'}^<(E)-t_{nn'}G_{n'n}^<(E)]
  \label{eq:current}
\end{equation}

$I_0$=2e/h = 77.48092 $\mu$A/eV is the natural unit of bond charge current density. The lesser
Green's function in the absence of interactions can be resolved exactly as
$G^<(E)=G^r(E)[\Gamma_Lf_L+\Gamma_Rf_R]G^a(E)$ where
$\Gamma_{L(R)}=i(\sum_{L(R)}-\sum_{L(R)}^{\dagger})$ is the left (right) contact broadening function
and $f_{L(R)}$ is the Fermi distribution of the left (right) contact.  $t_{n'n}$ is the hopping
parameter between sites $n'$ and $n$, in our BLG zigzag nanorribon represents  $\gamma_0$ for
intralayer bond  current and $\gamma_1$ for interlayer bond current.  In order to quantify the
electron flow in a layer we defined the layer current density as:

\begin{equation}
\label{eq:currentfield2}
I_m=\sum_{k \in m}J_{k},
\end{equation}

$k$ represents a site in the central region of the nanoribbon in layer $m=T,B$; 
$J_{k}=\sum_{n'}J_{kn'}$ is the total current at site $k$ calculated adding the bond current  
(\eqref{eq:current}) between site $k$ and its neighboring sites $n'$. Once again, since we are 
working on a
pristine nanoribbon the current density in any slide of our device is exactly the same, because of
that we associated it to a layer current density in \eqref{eq:currentfield2}.

Complementary to current density, charge density at site $k$ can also be expressed using
the lesser Green's function as:

\begin{equation}
\label{eq:cchargedensity}
\tilde{\rho}_c(k)=\frac{e}{2\pi i}\int_{E_F-eV/2}^{E_F + eV/2} dE{G_{k,k}^<(E)}.
\end{equation}

At equilibrium, all states are
occupied as specified by the Fermi-Dirac distribution ($f(E)$) and  the lesser Green's function
acquires the simple form $G^<(E)=if(E)A(E)$. Where $A(E)=i(G^r-G^a)$ is the spectral function, which
is related to the LDOS as $\rho(r,E) = \frac{1}{2\pi}A(r,E)$ \cite{Datta99}. 
It is noteworthy that at low bias and low temperature $\rho_c \approx e^2V\rho(E_F)$, and clearly 
it is 
observed that charge density ($\rho_c$) has the same local distribution of LDOS ($\rho$). Given we 
are interested in how charge and current distributions are related; with no loss of generality, to 
keep explanations and figures as simples as possible, we will refer from now on to LDOS as charge 
distribution.
To quantify and visualize how
charge (LDOS) is distributed over one layer it is defined the charge density per layer
as:

\begin{equation}
\label{eq:cchargedensity2}
\rho_m =\sum_{k \in m}\tilde{\rho}_k.
\end{equation}

The Green's function formalism has succeed in reproducing scanning probe microscopy 
experiments\cite{CreFG03,MetB05}
providing
a framework to interpret the measured charge map, electron flow as well as predicting new effects.

%================================================================
\section{Results for Charge and current density }
\label{sec:Charge and current density}
%================================================================

\begin{figure}[ht!]
\begin{center}
\includegraphics[width=0.945\columnwidth]{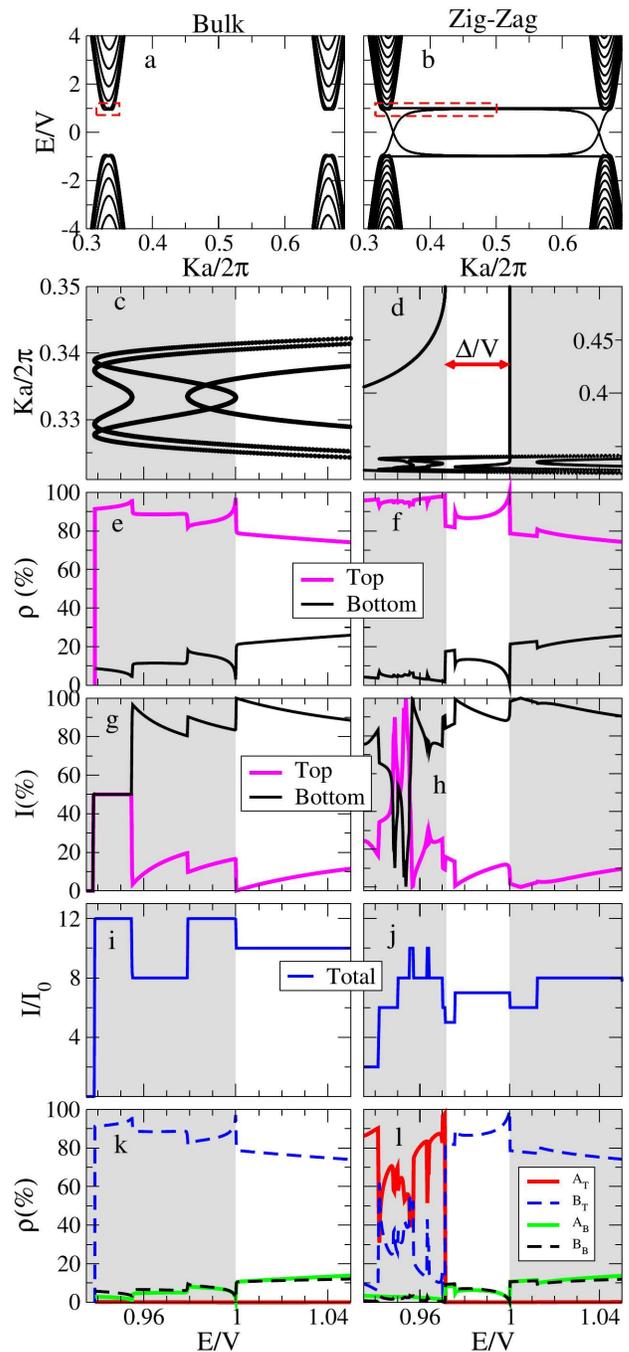}
%\vspace{1cm}
\caption{ (color online)
The left column shows results for a bulk biased BLG, while the right
column are for BLG nanoribbon with zigzag edges (for both cases a potential difference
V=0.07 eV and a width of $N=300$ atoms is considered.
(a)-(b) Band structures. (c)-(d) Zoom into the band structure's regions marked by the dashed lines
in
(a) and (b). $\Delta$ is the minimum separation between the flat and the dispersive edge state
bands
for the zigzag ribbon. (e)-(f) Percentage of the charge density in each layer.  (g)-(h) Current
density on each layer.  (i)-(j) Total current density on the BLG.
(k)-(l) The contribution of each sublattice to the charge density.}
\label{fig:Fig2}
\end{center}
\end{figure}

In \Fref{fig:Fig2} we show the band structure and details of the charge and the current densities
for BLG under
the influence of an applied voltage difference of $2V=0.14$ eV between the two layers. The right
column shows
the results for a BLG nanoribbon with zigzag edges and width of $N=300$ atoms, while the results at
the left
column are for a bulk BLG (for which the same width of $N=300$ atoms was considered with periodical
boundary conditions).

\subsection{Band Structures}

The band structure for the bulk BLG in \Fref{fig:Fig2}(a) evidences the opening of the energy gap of
approximately 2V (observe that the energy scale is normalized by the bias voltage V).
The presence of the zigzag edges introduces edge states in the gap region of the band structure, as
observed
in \Fref{fig:Fig2}(b). Zooms into the band structure's regions marked by the dashed lines in
\Fref{fig:Fig2}(a) and (b) are shown in \Fref{fig:Fig2}(c) and (d), respectively.
For the bulk, we see in \Fref{fig:Fig2}(c) the well-known ``Mexican-hat" structure, due to the
applied bias,
mixed to other higher bands for this system size and bias (the wider the nanoribbon considered the
higher is the
density of bands and the band mixing in this region). For the zigzag case, one can see in more
detail in \Fref{fig:Fig2}(d)
that in addition to the usual band structure this region contains two edge states energy bands: a
flat band at $E/V=1$ and
a dispersive band for $E/V \leq 1 - \Delta/V$, in agreement with previous works
\cite{CasPL08,CasNM10,YaoYN09,LiZX10,RhiK08}. The minimum
separation between the edge states bands is $\Delta$=2V$\frac{\gamma_1^2}{\gamma_1^2+\gamma_0^2}$
(this expression is derived
from the difference between dispersive band  and flat band \cite{CasPL08} at $ka/2\pi=0.5$).
Observe that $\Delta$ increases linearly with the external
bias V and does
not depend on the nanoribbon width.

\subsection{Charge and Current Density in each Layer}

In \Fref{fig:Fig2}(e) and (f) we show the percentage of the total charge density of
the bilayer which is accumulated in each of the layers (top or bottom) - once charge 
in each layer is obtained from \eqref{eq:cchargedensity2}, its proportion
with respect to the total charge in both layers is calculated .
Both for the system with zigzag edges and for the bulk there is a clear unbalance 
between layers, with  electronic charge density being
concentrated predominantly (from 75 to 100$\%$) over the top layer for the entire energy range
shown. 

The percentage of the current density  over each layer is calculated in the same way 
from \eqref{eq:currentfield2} and the results are shown in  Fig.
\ref{fig:Fig2}(g) and (h), for bulk and zigzag, respectively. Comparing charge and
current densities in each layer, one can see that although the charge densities are highly
concentrated over the top layer, the current
densities are higher on the bottom layer for a wide energy range, both
for bulk or zigzag BLG. This behavior
is counterintuitive and contradicts the most basic theoretical model of charge flow.

\Fref{fig:Fig2}(i) and (j) show the total current density ($I$ divided by $I_0$), which corresponds
to the summation of the currents in the top and the bottom layers. The total current density is
directly proportional to the conductance of the system.

\subsection{The Role of the Sublattices}

To investigate the origin of the discrepancy between the charge density and current density in each
layer, we
compute separately the contribution of each sublattice to the charge density, as shown in
\Fref{fig:Fig2}(k) and (l).
This gives us an important clue to understand the phenomenon: the charge is not only predominantly
over one layer
(the top layer), there is also a sublattice polarization in this layer.

For the bulk, \Fref{fig:Fig2}(k), we observe that the charge on the top layer is entirely over
the sublattice $B_T$,
while sublattice $A_T$ shows zero contribution to the charge density in the entire energy range
shown.
This effect comes from the sublattice asymmetry introduced by the AB-stacking in BLG\cite{CasNM10}:
the dimer
sites ($A_T$ and $B_B$)
hybridized their orbitals to form higher energy bands, being the charge density for low energy
states
located mostly on non-dimer sites  ($B_T$ and $A_B$) \cite{McC06,KosA06}.  Here we show in
\Fref{fig:Fig2}(k) how this
sublattice asymmetry is preserved in the top layer after the application of the voltage difference
between the layers. We
see that although the charge on top layer keeps completely located over only one sublattice 
( non-dimer $B_T$),
the charge density over
the bottom layer is mostly sublattice unpolarized, ie, equally shared between the 
two sublattices. This interesting characteristic of these systems, which has 
already been point out in previous experimental\cite{KimKW13}, 
analytical\cite{McCK13,CasNM10} and numerical\cite{RamNT11} works, can possibly explain why 
the
current goes preferentially
over the bottom layer.

% \cite{YaoYN09}  - > checar esta referencia

For the zigzag nanoribbon, the density distribution in each sublattice becomes even more
interesting, \Fref{fig:Fig2}(l),
with a clear competition between the bulk effect in the AB-stacking just described and the
additional sublattice
polarization that is well known to occur around the zigzag edges \cite{CasPL08}, as 
we will show.
For energies $E/V > 1 - \Delta/V$, we see from \Fref{fig:Fig2}(l) that the sublattice distribution
is similar to that
described to the bulk, with charge on top layer only over $B_T$ sublattice.
However, for energies $E/V \leq 1 - \Delta/V$, while \Fref{fig:Fig2}(f) tells us that the charge
is still over the top layer (more than $90\%$),  \Fref{fig:Fig2}(l) shows that there is an
inversion in the subltattice:
$A_T$ sublattice is now predominant, with some oscillations. Comparing to the band structure in
\Fref{fig:Fig2}(d), we see
 that in this region the dispersive edge-state band plays an important role.
The energy split of size $\Delta$ corresponds in fact to the split of states localized on opposite
edges of the
top layer.  The states from the flat band at  $E/V=1$ are
located on the edge of the top layer where the outermost atoms are $B_T$ atomic sites, the same
sublattice that is privileged by the
AB-stacking. On the other hand,  as in zigzag graphene nanoribbons
the outermost atoms in opposite edges belong to different sublattices, the states from
the dispersive band at $E/V \leq 1 - \Delta/V$ are located on the edge where the outermost atoms of
the top layer are $A_T$ sites.
And here is where the competition between edge and bulk arises, leading to the oscillations between
sublattices observed, with an advantage
to the  $A_T$ sites, i.e., the edge state localization effect being more robust than the bulk
effect.

%============================================================================
\section{Mapping the Spatial Distribution of Charge and Current over the BLG}
%============================================================================

\begin{figure*}[t!]
\begin{center}
\includegraphics{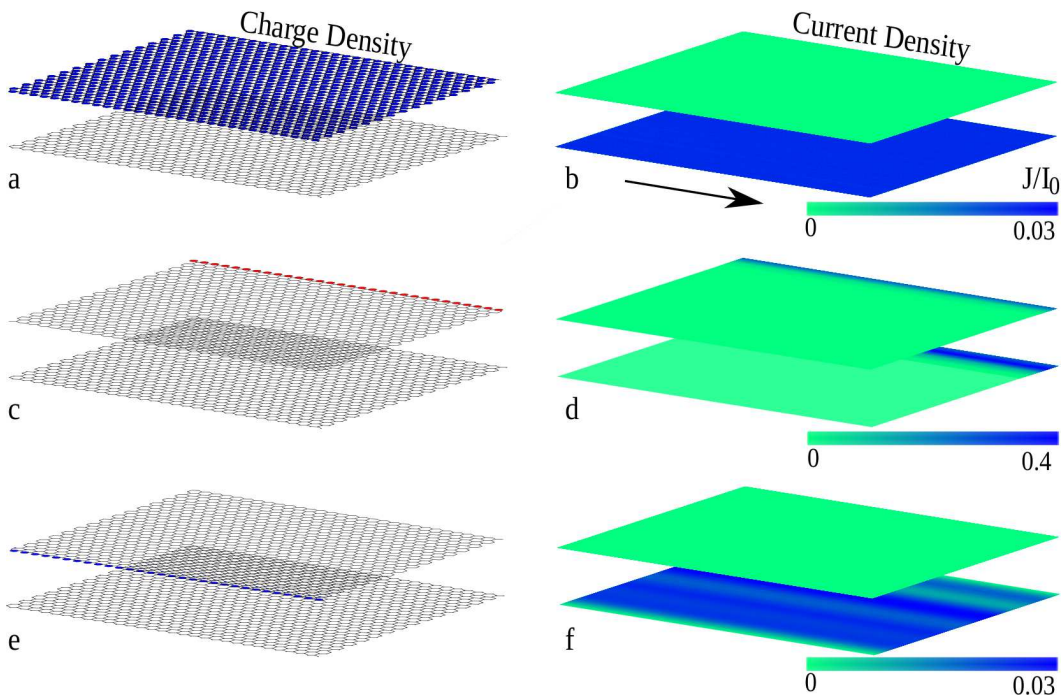}
\caption{ (color online) Spatial distribution of charge densities
(left) and current densities (right) over each layer of the BLG.
Results for a bulk system (periodical boundary conditions) are shown in (a)-(b) for N=300 and
E/V=1.003.  Results for a BLG
nanoribbon with zigzag edges and $N=300$ are shown in (c)-(d) for E/V=0.0965; and in (e)-(f) for
E/V=1.003 (e)-(f).
The charge densities are schematically represented here for a narrower nanoribbon, where the radius
of each
circle is proportional to the amplitude of charge density and different colors stand for different
sublattices.
The current densities are evaluated at
different sites using \eqref{eq:current}.}
\label{fig:Fig3}
\end{center}
\end{figure*}

In a lattice, we can imagine one electron injected from the left contact
hopping from site to site until reaching the right contact. Clearly one electron is enabled to hop
on its nearest neighbor if there are electronic states available there; in this regard the spatial
distribution of charge and current densities are expected to be related to each other.

In the discussion of the previous section we have already identified that the polarization of the
charge density to
only one of the sublattices plays an important role in the discrepancies observed between charge
and
current densities
in each layer. Here, \Fref{fig:Fig3} helps us to observe in more detail the spatial distribution
of charge
and current
densities over each layer (and each sublattice) of the bilayer systems.
The systems considered in this calculation of \Fref{fig:Fig3} are exactly the same from
\Fref{fig:Fig2}: biased BLG of
300 carbon atoms in width,
with periodical boundary conditions for the bulk and zigzag edges for the nanoribbon, V=0.07eV.
For the representations of the spatial distribution of the charge density, the density on each
atomic
site is shown here as proportional to the radius of the disk and its
color stands for sublattice: red for $A_T$ and $A_B$ and blue for $B_T$ and $B_B$ (same color
scheme
shown in
\Fref{fig:device}(b)).

Initially, in \Fref{fig:Fig3}(a) and (b) we map charge and current spatial density for a
 bulk BLG, avoiding in this case any complication introduced by the edge states.
This distribution corresponds to the energy $E/V=1.003$ - at this energy, Fig2(e) tells us that
80\%
of the charge is located on top layer, while Fig.2(g) shows that current density is much higher in
bottom layer and nearly zero on top layer.
\Fref{fig:Fig3}(a) shows that on the top layer charge is completely located on
non-dimer $B_T$ sites, being homogeneously distributed over the layer.
 On the bottom layer, although it is not appreciated in
\Fref{fig:Fig3}(a) because $\rho_B$ is approximately 10 times smaller than $\rho_T$; there is a
homogeneous charge density on $A_B$ and $B_B$ sites (9\% each, see \Fref{fig:Fig2}(k)). When
current density is calculated, as shown in \Fref{fig:Fig3}(b), it is appreciated that current is
homogeneously distributed over  bottom layer while is nearly zero over top layer. This can be
understood on account of charge is completely localized on non-dimer sites ($B_B$). Electrons on
these sites can not jump on its nearest neighbors, causing no electron flow over top layer. On
the bottom layer, in spite of $\rho_B < \rho_T$, charge is homogeneously distributed over both
sublattices, allowing electron hopping among sites.

For the BLG zigzag nanoribbons, bias lifts edge states degeneracy.
We see from \Fref{fig:Fig3}(c) and (e) that the charge densities for energies corresponding
to the split edge state bands, $E/V=0.965 \approx 1-\Delta/V$ and $E/V=1.003 \approx 1$, are highly
localized on opposite zigzag edges of the top layer, in agreement to previous
calculations \cite{CasNM10,YaoYN09,RhiK08,LiZX10}. For the first energy one can see 
from \Fref{fig:Fig2}(l) that nearly
80\% of the charge density is located on dimer sites $A_T$ while
18\% of the charge is located on non-dimer sites $B_T$. Spatial mapping of the charge density
reveals in \Fref{fig:Fig3}(c) an edge state located on only one of the edges of the top layer:
the edge whose outermost atoms are from sublattice $A_T$. 
Once again due to the considerable difference between the densities in the two sublattices, only
the edge state is appreciable. However, charge density is also homogeneously distributed on 
non-dimer
sites ($B_T$) of the top layer. Overlapping  of the exponentially decaying edge state ($A_T$) and
the homogeneously distributed state ($B_T$)
 creates a high current density on this edge of the top layer, as depicted on Fig.
\ref{fig:Fig3}(d).  This figure also shows a high current density on the bottom layer 
right bellow this edge, its
origin is similar to the top layer current: this edge at bottom layer terminates at a non-dimer 
sites $A_B$
sustaining edge states while dimer sites $B_B$  have an enhanced charge density caused by the top
layer edge state; these two states overlap creating the highly charge current observed. When  next
to nearest neighbors are included in monolayer zigzag nanoribbon edge states acquire velocity, this
however does not affect charge or current density. 

The effect of the sublattice symmetry breaking is also observed for $E/V=1.003 \approx 1$. At this
energy, edge states localize on the other edge, the one whose outermost atoms are from $B_T$ 
sublattice, 
as shown in \Fref{fig:Fig3}(e). Considering that this sublattice $B_T$ corresponds to non-dimer 
atoms, 
current on bottom layer is not affected for this energy, as shown in \Fref{fig:Fig3}(f):
current is distributed over the whole layer. Over the top layer there is no current because charge 
is
completely localized on $B_T$ sites, this situation is reminiscent of bulk biased BLG.

%================================================================
\section{Current dependence on bias voltage and size}
%================================================================

\begin{figure}[t!]
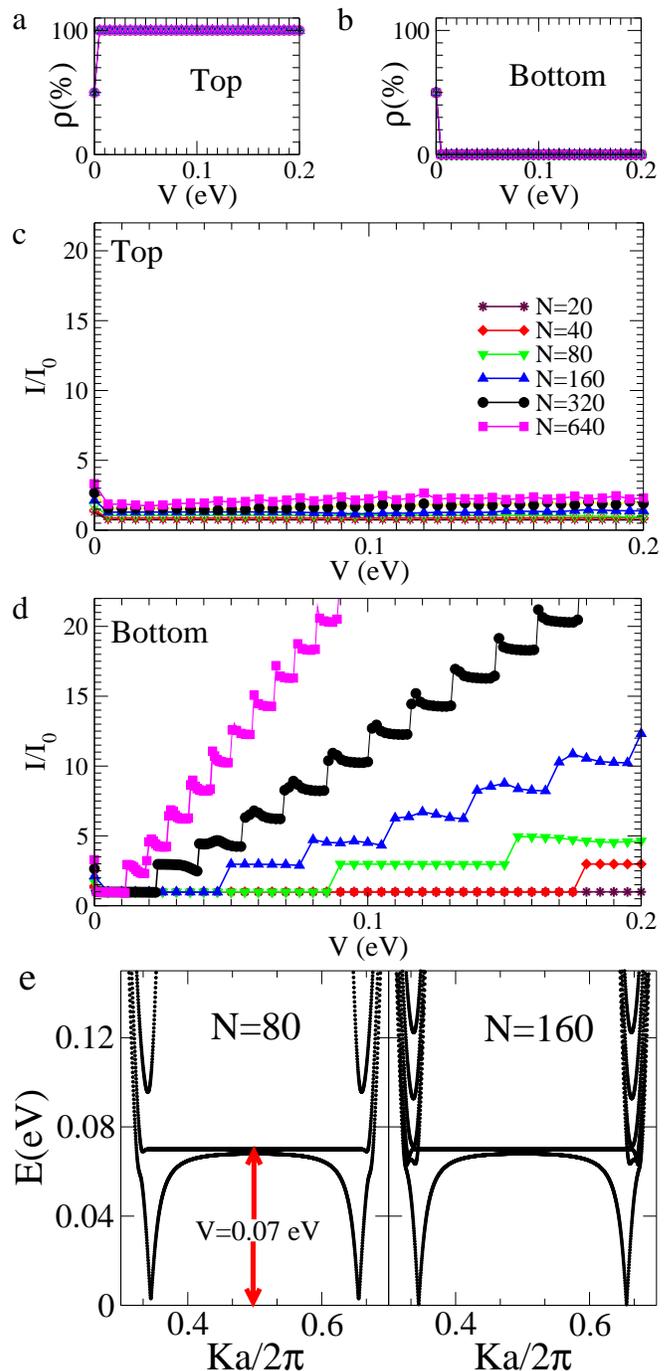

\begin{center}
\includegraphics[width=1\columnwidth]{Fig4a.eps}
\includegraphics[width=1\columnwidth]{Fig4b.eps}
\caption{(color online) (a)-(b) Percentage of charge density on top and bottom layer as a function
of the bias
$V$ applied between the layers, calculated for $E/V=1$. (c)-(d) Total current density on top and
bottom layer, as a function of bias V for $E/V=1$ .
Several widths of zigzag biased BLG are shown: from $N=20$ to $N=640$. (e)  Band structure for a
BLG nanoribbon with zigzag edges, $V=0.07$ eV, for two different sizes: $N=80$  and  $N=160$.}
\label{fig:Fig4}
\end{center}
\end{figure}

In this section we focus on the effects of the bias voltage strength and the width of the
nanoribbons on the current and charge densities. For this purpose,
we need first to choose a fixed energy.
The experimental observation of a current density highly localized on one of the edges
of a BLG nanoribbon, like the current shown in \Fref{fig:Fig3}(d), would require an
extremely clean sample, as edge disorder would scatter electrons, degrading the current
and destroying its spatial localization \cite{MucCL09,CreR09,MucL10}.
On the other hand,  setting $E/V \approx \pm1$  for a BLG nanorribon offers
control of  the layer pseudospin: the layer in which charge current is conducted (top or bottom),
in a similar way observed for bulk systems (as observed for the current in
Figs. \ref{fig:Fig2} and \ref{fig:Fig3}) and avoiding the edge disorder sensitivity.
Therefore, we choose to investigate here how the current and the charge densities
vary with system size and with bias voltage at energy $E/V =1$: results are shown
in \Fref{fig:Fig4}.  Figures \ref{fig:Fig4}(a) and (b) show that for wider nanoribbons,
bias voltage variation and ribbon width do not modify the complete charge density polarization
on the top layer.  As we have seen, for this energy, charge mostly localizes on
non-dimer sites of top layer reducing nearly to zero the current density over top layer. Here we
show
in Figs. \ref{fig:Fig4}(c) that this characteristic is maintained with increasing bias and system
sizes.
\Fref{fig:Fig4}(d) shows a  different evolution for the current density on the bottom layer:
for wider nanoribbons or larger bias voltages, the current density rises in steps.
This dependence is understood from the band
structure of the biased BLG nanoribbon,  as seen in \Fref{fig:Fig4}(e) for $V=0.07$eV and
widths corresponding to $N=80$ and $N=160$.
It is appreciated that flat bands are fixed at $E/V=1$ and do not depend on the nanoribbon width.
On the other hand, the number of dispersive bands around $E/V=1$ increases with N, adding more
conducting channels. For that reason current density
evolves in a plateau-like structure. The peaks observed, at the beginning of each plateau, for wider
nanoribbons $N=320$ and $N=640$ are created by the ``Mexican-hat'' structure of bands crossing
$E/V=1$.

%================================================================
\section{EFFECTS OF DISORDER AND NEXT-NEAREST HOPPINGS} 
\label{Disorder-nnn}
%================================================================

In this section we discuss how disorder and next-nearest neighbor hoppings affect the picture 
presented in the previous sections. Mainly, we show here that although there are important 
features introduced by disorder and by further hoppings in the model, the assymmetry between charge 
and current density distributions is still present, therefore, the effects previously discussed are 
robust. 

\begin{figure}[t!]
\begin{center}
\includegraphics[width=1\columnwidth]{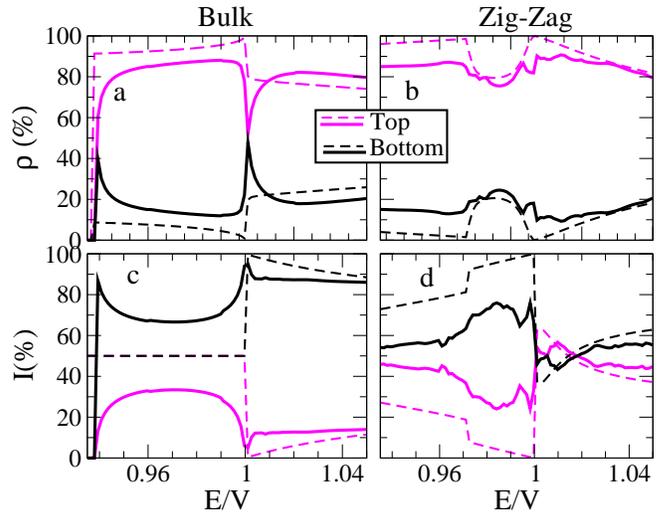}
\caption{(color online) Dashed lines are for non-disordered systems, while solid lines show the 
effects of the inclusion of on-site correlated disorder ($W/\gamma_0=0.5$) in the biased BLG 
(V=0.07 eV). Systems considered here have 80x40 sites in each layer. The left column shows results 
for a bulk, while the right column is for BLG nanoribbon with zigzag edges. (a)-(b) Percentage of 
the charge density in each layer. (c)-(d) Current density on each layer.}
\label{fig:Fig5}
\end{center}
\end{figure}

In \Fref{fig:Fig5} we show the comparison between a non-disorderd BLG (dashed lines) and a 
disordered system (solid lines), again analysing both for a bulk and for a zig-zag nanoribbon. Each 
layer of these systems here have width of N = 80 atoms and 40 atoms in the length between the 
contacts. To account for disorder, we introduce a Gaussian-correlated on-site disorder for each 
layer\cite{PerS08b}, with site energies ramdomly sorted in a range of width $W/\gamma_0= 0.5$ and 
correlation 
length $\lambda=2a$ for bottom layer and $\lambda=5a$ for top layer ($a = 2.46 \mathring{A}$). 
Larger 
correlation length for the top layer is due to its higher distance from the substrate. For the 
charge density distribution, \Fref{fig:Fig5}(a) and (b), we see that the disorder does not alter 
significantly the clear concentration of the charge on the top layer for all the energy range shown 
(except for the bulk at energy exactly E/V=1). For the current density, one can see that, for 
the bulk, \Fref{fig:Fig5}(c), disorder not only does not alter the fact that about 80\% of the 
current goes through bottom layer for $E/V>1$ but also disorder destroys the equilibrium of the 
currents that appears at this system size  for $E/V<1$, producing again a predominance of the 
current 
over bottom layer and an unbalance between charge and current in the two layers. In the presence of 
edges, for the zigzag case shown in \Fref{fig:Fig5}(d), we see that although the current in the 
disordered system still flows predominantly through the bottom layer, the percentages are smaller 
than in the non-disordered. Edge states are probably the most affected by the disorder, however 
further investigations would be necessary to clarify the role of disorder separately on edge and 
bulk current states.

\begin{figure}[t!]
\begin{center}
\includegraphics[width=1\columnwidth]{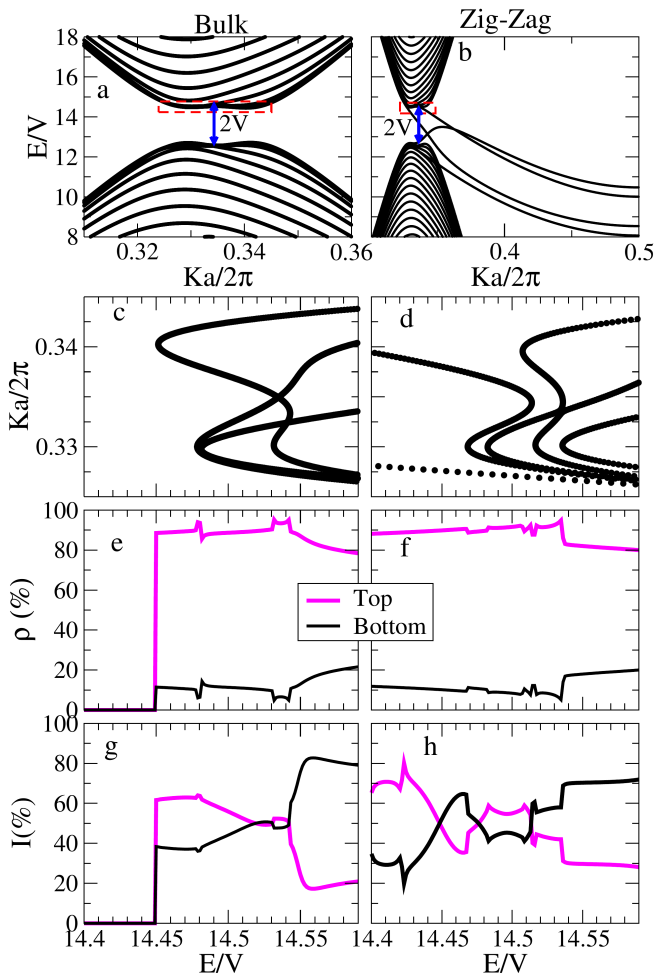}
\caption{(color online) Analysis with the inclusion of next-nearest neighbor hoppings in the 
model for the biased BLG. The left column is for a bulk, while the right column is for BLG 
nanoribbon with zigzag edges (for both cases V=0.07 eV and a width of N = 300 atoms is considered). 
(a)-(b) Band structures. (c)-(d) Zoom into the band structure’s regions marked by the dashed lines 
in (a) and (b). (e)-(f) Percentage of the charge density in each layer. (g)-(h) Current density on 
each layer.}
\label{fig:Fig6}
\end{center}
\end{figure}

In \Fref{fig:Fig6} we turn our attention to the effects of the inclusion of further hoppings in the 
tight-binding model. These results are for a non-disordered BLG system of the same size (N=300) and 
same bias voltage (V=0.07 eV) considered in \Fref{fig:Fig2}. The difference is that now we include 
next-nearest neighbors in each layer ($\gamma_2=0.316$eV) and also two extra interlayer coupling 
parameters: $\gamma_3 = 0.38$eV, the interlayer coupling between non-dimer sites $A_B$ and $B_T$, 
and 
$\gamma_4 = 0.14$eV, the interlayer coupling between dimer and non-dimer sites $A_T$ and $A_B$, or 
$B_T$ 
and $B_B$ \cite{McCK13,KuzCV09}. These induce a trigonal warping and give rise to electron-hole 
asymmetry\cite{McCK13,CasGPN09}, as observed in\Fref{fig:Fig6}(a) and (b). For the bulk, we see 
once 
again the opening of the energy gap of approximately 2V, due to the applied bias. The band 
structure 
is not anymore symmetrical around E/V=0, there is an energy shift of the band structure to around 
$E/V\approx13.5$ \cite{NilCG08}. We can see the mixing of higher bands for this system size and 
bias. For the zigzag case, one can see that edge states acquire velocity\cite{SasMS06}. 
\Fref{fig:Fig6} (c) and (d) are zooms into the band structure’s regions marked by the dashed lines.

In \Fref{fig:Fig6}(e) and (f) we show the percentage of the total charge density of the bilayer 
which is accumulated in each of the layers (top or bottom). Comparing them to \Fref{fig:Fig2}(e) 
and (f), one observe that both for the bulk system and for the one with zigzag edges, the further 
hoppings here do not affect at all the polarization of charge towards the top layer. The 
distribution of the current density in each layer is shown in \Fref{fig:Fig6}(g) and (h), for bulk 
and zigzag, respectively. 
One can see now that, in general, the current is not as polarized toward the bottom layer as 
it 
is for the nearest neighbor hopping seen in \Fref{fig:Fig2}(g) and (h). Nevertheless, the unbalance 
between charge and current densities is still clear, and even considering the non-vertical 
interlayer couplings and the next-nearest intralayer hoppings, there are clear energy regions where 
the current flows more throughout the bottom layer. There are also switches to energy regions where 
the current is shared between both layers or is predominant over top layer. Further investigations 
are important here to elucidate the exact mechanisms causing these switches. 

\section{Conclusions}

An external electric field perpendicular to a Bernal-stacked bilayer graphene, breaks layer and 
sublattice
symmetry; localizing charge over two sublattices in different layers. Under this charge landscape,
it is not obvious how charge current flows within the sample. 
We have shown here that
current  distribution is highly affected by the polarization of charge to only one sublattice, as 
well as 
by the geometry of the system (width considered and presence or not of 
edges) and by the strength of the electric field.

We demonstrate that current does not necessarily flow over regions of the system with higher charge 
density, even when next-nearest neighbor hoppings are included in our model. For some energy 
ranges, 
charge can be polarized to one layer, while the current is equally distributed over both layers. 
There are also considerable energy ranges for which the current flows predominantly over the layer 
with much lower charge density. We show that this effect can be explained by the sublattice 
polarization of charge in the AB-stacking biased BLG, and that it is robust against disorder. 
Therefore, to design applications of bilayer graphene in digital electronics, it is essential to 
calculate not only the charge distribution in each layer, but also the current density distribution 
in each layer, as it presents much richer details than the more monotonic behavior of the charge 
distribution. 

\begin{acknowledgments}
The authors are grateful to P. A. Schulz for a critical reading of the manuscript. C.J.P.\
acknowledges financial support from FAPESP (Brazil), A.L.C.P.\ acknowledges partial support
from CNPq (Brazil). Part of the numerical simulations were performed at the computational facilities
from CENAPAD-SP, at Campinas State University.
\end{acknowledgments}

%merlin.mbs apsrev4-1.bst 2010-07-25 4.21a (PWD, AO, DPC) hacked
%Control: key (0)
%Control: author (8) initials jnrlst
%Control: editor formatted (1) identically to author
%Control: production of article title (-1) disabled
%Control: page (0) single
%Control: year (1) truncated
%Control: production of eprint (0) enabled

%

%\bibliography{/home/carlos/Documentos/bibliography/bibliograph}
%\bibliography{/home/carlos/Documentos/bibliography/apsrev4-1}

\begin{thebibliography}{47}%
\makeatletter
\providecommand \@ifxundefined [1]{%
 \@ifx{#1\undefined}
}%
\providecommand \@ifnum [1]{%
 \ifnum #1\expandafter \@firstoftwo
 \else \expandafter \@secondoftwo
 \fi
}%
\providecommand \@ifx [1]{%
 \ifx #1\expandafter \@firstoftwo
 \else \expandafter \@secondoftwo
 \fi
}%
\providecommand \natexlab [1]{#1}%
\providecommand \enquote  [1]{``#1''}%
\providecommand \bibnamefont  [1]{#1}%
\providecommand \bibfnamefont [1]{#1}%
\providecommand \citenamefont [1]{#1}%
\providecommand \href@noop [0]{\@secondoftwo}%
\providecommand \href [0]{\begingroup \@sanitize@url \@href}%
\providecommand \@href[1]{\@@startlink{#1}\@@href}%
\providecommand \@@href[1]{\endgroup#1\@@endlink}%
\providecommand \@sanitize@url [0]{\catcode `\\12\catcode `\$12\catcode
  `\&12\catcode `\#12\catcode `\^12\catcode `\_12\catcode `\%12\relax}%
\providecommand \@@startlink[1]{}%
\providecommand \@@endlink[0]{}%
\providecommand \url  [0]{\begingroup\@sanitize@url \@url }%
\providecommand \@url [1]{\endgroup\@href {#1}{\urlprefix }}%
\providecommand \urlprefix  [0]{URL }%
\providecommand \Eprint [0]{\href }%
\providecommand \doibase [0]{http://dx.doi.org/}%
\providecommand \selectlanguage [0]{\@gobble}%
\providecommand \bibinfo  [0]{\@secondoftwo}%
\providecommand \bibfield  [0]{\@secondoftwo}%
\providecommand \translation [1]{[#1]}%
\providecommand \BibitemOpen [0]{}%
\providecommand \bibitemStop [0]{}%
\providecommand \bibitemNoStop [0]{.\EOS\space}%
\providecommand \EOS [0]{\spacefactor3000\relax}%
\providecommand \BibitemShut  [1]{\csname bibitem#1\endcsname}%
\let\auto@bib@innerbib\@empty
%</preamble>
\bibitem [{\citenamefont {Topinka}\ \emph {et~al.}(2000)\citenamefont
  {Topinka}, \citenamefont {LeRoy}, \citenamefont {Shaw}, \citenamefont
  {Heller}, \citenamefont {Westervelt}, \citenamefont {Maranowski},\ and\
  \citenamefont {Gossard}}]{TopLS00}%
  \BibitemOpen
  \bibfield  {author} {\bibinfo {author} {\bibfnamefont {M.~A.}\ \bibnamefont
  {Topinka}}, \bibinfo {author} {\bibfnamefont {B.~J.}\ \bibnamefont {LeRoy}},
  \bibinfo {author} {\bibfnamefont {S.~E.~J.}\ \bibnamefont {Shaw}}, \bibinfo
  {author} {\bibfnamefont {E.~J.}\ \bibnamefont {Heller}}, \bibinfo {author}
  {\bibfnamefont {R.~M.}\ \bibnamefont {Westervelt}}, \bibinfo {author}
  {\bibfnamefont {K.~D.}\ \bibnamefont {Maranowski}}, \ and\ \bibinfo {author}
  {\bibfnamefont {A.~C.}\ \bibnamefont {Gossard}},\ }\href {\doibase
  10.1126/science.289.5488.2323} {\bibfield  {journal} {\bibinfo  {journal}
  {Science}\ }\textbf {\bibinfo {volume} {289}},\ \bibinfo {pages} {2323}
  (\bibinfo {year} {2000})},\ \Eprint
  {http://arxiv.org/abs/http://www.sciencemag.org/content/289/5488/2323.full.pdf}
  {http://www.sciencemag.org/content/289/5488/2323.full.pdf} \BibitemShut
  {NoStop}%
\bibitem [{\citenamefont {Topinka}\ \emph {et~al.}(2001)\citenamefont
  {Topinka}, \citenamefont {LeRoy}, \citenamefont {Westervelt}, \citenamefont
  {Fleischmann}, \citenamefont {Heller},\ and\ \citenamefont
  {Maranowski}}]{TopLW01}%
  \BibitemOpen
  \bibfield  {author} {\bibinfo {author} {\bibfnamefont {M.~A.}\ \bibnamefont
  {Topinka}}, \bibinfo {author} {\bibfnamefont {B.~J.}\ \bibnamefont {LeRoy}},
  \bibinfo {author} {\bibfnamefont {S.~E.~J.}\ \bibnamefont {Westervelt},
  \bibfnamefont {R.~M.and~Shaw}}, \bibinfo {author} {\bibfnamefont
  {R.}~\bibnamefont {Fleischmann}}, \bibinfo {author} {\bibfnamefont {E.~J.}\
  \bibnamefont {Heller}}, \ and\ \bibinfo {author} {\bibfnamefont {A.~C.}\
  \bibnamefont {Maranowski}, \bibfnamefont {K.~D.and~Gossard}},\ }\href
  {\doibase 0.1038/35065553} {\bibfield  {journal} {\bibinfo  {journal}
  {Nature}\ }\textbf {\bibinfo {volume} {410}},\ \bibinfo {pages} {183}
  (\bibinfo {year} {2001})}\BibitemShut {NoStop}%
\bibitem [{\citenamefont {{A. H. Castro Neto}}\ \emph
  {et~al.}(2009)\citenamefont {{A. H. Castro Neto}}, \citenamefont {Guinea},
  \citenamefont {Peres}, \citenamefont {Novoselov},\ and\ \citenamefont
  {Geim}}]{CasGPN09}%
  \BibitemOpen
  \bibfield  {author} {\bibinfo {author} {\bibnamefont {{A. H. Castro Neto}}},
  \bibinfo {author} {\bibfnamefont {F.}~\bibnamefont {Guinea}}, \bibinfo
  {author} {\bibfnamefont {N.~M.~R.}\ \bibnamefont {Peres}}, \bibinfo {author}
  {\bibfnamefont {K.~S.}\ \bibnamefont {Novoselov}}, \ and\ \bibinfo {author}
  {\bibfnamefont {A.~K.}\ \bibnamefont {Geim}},\ }\href {\doibase
  10.1103/RevModPhys.81.109} {\bibfield  {journal} {\bibinfo  {journal} {Rev.
  Mod. Phys.}\ }\textbf {\bibinfo {volume} {81}},\ \bibinfo {eid} {109}
  (\bibinfo {year} {2009})}\BibitemShut {NoStop}%
\bibitem [{\citenamefont {Rycerz}\ \emph {et~al.}(2007)\citenamefont {Rycerz},
  \citenamefont {Tworzydlo},\ and\ \citenamefont {Beenakker}}]{RycTB07}%
  \BibitemOpen
  \bibfield  {author} {\bibinfo {author} {\bibfnamefont {A.}~\bibnamefont
  {Rycerz}}, \bibinfo {author} {\bibfnamefont {J.}~\bibnamefont {Tworzydlo}}, \
  and\ \bibinfo {author} {\bibfnamefont {C.~W.~J.}\ \bibnamefont {Beenakker}},\
  }\href {\doibase 10.1126/science.1065389} {\bibfield  {journal} {\bibinfo
  {journal} {Nat Phys}\ }\textbf {\bibinfo {volume} {3}},\ \bibinfo {pages}
  {172} (\bibinfo {year} {2007})}\BibitemShut {NoStop}%
\bibitem [{\citenamefont {San-Jose}\ \emph {et~al.}(2009)\citenamefont
  {San-Jose}, \citenamefont {Prada}, \citenamefont {McCann},\ and\
  \citenamefont {Schomerus}}]{SanPM09}%
  \BibitemOpen
  \bibfield  {author} {\bibinfo {author} {\bibfnamefont {P.}~\bibnamefont
  {San-Jose}}, \bibinfo {author} {\bibfnamefont {E.}~\bibnamefont {Prada}},
  \bibinfo {author} {\bibfnamefont {E.}~\bibnamefont {McCann}}, \ and\ \bibinfo
  {author} {\bibfnamefont {H.}~\bibnamefont {Schomerus}},\ }\href {\doibase
  10.1103/PhysRevLett.102.247204} {\bibfield  {journal} {\bibinfo  {journal}
  {Phys. Rev. Lett.}\ }\textbf {\bibinfo {volume} {102}},\ \bibinfo {pages}
  {247204} (\bibinfo {year} {2009})}\BibitemShut {NoStop}%
\bibitem [{\citenamefont {Akhmerov}\ and\ \citenamefont
  {Beenakker}(2007)}]{AkhB07}%
  \BibitemOpen
  \bibfield  {author} {\bibinfo {author} {\bibfnamefont {A.~R.}\ \bibnamefont
  {Akhmerov}}\ and\ \bibinfo {author} {\bibfnamefont {C.~W.~J.}\ \bibnamefont
  {Beenakker}},\ }\href {\doibase 10.1103/PhysRevLett.98.157003} {\bibfield
  {journal} {\bibinfo  {journal} {Phys. Rev. Lett.}\ }\textbf {\bibinfo
  {volume} {98}},\ \bibinfo {pages} {157003} (\bibinfo {year}
  {2007})}\BibitemShut {NoStop}%
\bibitem [{\citenamefont {Gunlycke}\ and\ \citenamefont
  {White}(2011)}]{GunW11}%
  \BibitemOpen
  \bibfield  {author} {\bibinfo {author} {\bibfnamefont {D.}~\bibnamefont
  {Gunlycke}}\ and\ \bibinfo {author} {\bibfnamefont {C.~T.}\ \bibnamefont
  {White}},\ }\href {\doibase \approx10.1103/PhysRevLett.106.136806} {\bibfield
   {journal} {\bibinfo  {journal} {Phys. Rev. Lett.}\ }\textbf {\bibinfo
  {volume} {106}},\ \bibinfo {pages} {136806} (\bibinfo {year}
  {2011})}\BibitemShut {NoStop}%
\bibitem [{\citenamefont {Schwierz}(2010)}]{Sch10}%
  \BibitemOpen
  \bibfield  {author} {\bibinfo {author} {\bibfnamefont {F.}~\bibnamefont
  {Schwierz}},\ }\href {\doibase 10.1038/nnano.2010.89} {\bibfield  {journal}
  {\bibinfo  {journal} {Nat Nano}\ }\textbf {\bibinfo {volume} {5}},\ \bibinfo
  {pages} {487} (\bibinfo {year} {2010})}\BibitemShut {NoStop}%
\bibitem [{\citenamefont {Pesin}\ and\ \citenamefont
  {MacDonald}(2012)}]{PesM12}%
  \BibitemOpen
  \bibfield  {author} {\bibinfo {author} {\bibfnamefont {D.}~\bibnamefont
  {Pesin}}\ and\ \bibinfo {author} {\bibfnamefont {A.~H.}\ \bibnamefont
  {MacDonald}},\ }\href {http://dx.doi.org/10.1038/nmat3305} {\bibfield
  {journal} {\bibinfo  {journal} {Nat Mater}\ }\textbf {\bibinfo {volume}
  {11}},\ \bibinfo {pages} {409} (\bibinfo {year} {2012})}\BibitemShut
  {NoStop}%
\bibitem [{\citenamefont {Castro}\ \emph {et~al.}(2008)\citenamefont {Castro},
  \citenamefont {Peres}, \citenamefont {Lopes~dos Santos}, \citenamefont
  {Neto},\ and\ \citenamefont {Guinea}}]{CasPL08}%
  \BibitemOpen
  \bibfield  {author} {\bibinfo {author} {\bibfnamefont {E.~V.}\ \bibnamefont
  {Castro}}, \bibinfo {author} {\bibfnamefont {N.~M.~R.}\ \bibnamefont
  {Peres}}, \bibinfo {author} {\bibfnamefont {J.~M.~B.}\ \bibnamefont
  {Lopes~dos Santos}}, \bibinfo {author} {\bibfnamefont {A.~H.~C.}\
  \bibnamefont {Neto}}, \ and\ \bibinfo {author} {\bibfnamefont
  {F.}~\bibnamefont {Guinea}},\ }\href {\doibase
  10.1103/PhysRevLett.100.026802} {\bibfield  {journal} {\bibinfo  {journal}
  {Phys. Rev. Lett.}\ }\textbf {\bibinfo {volume} {100}},\ \bibinfo {pages}
  {026802} (\bibinfo {year} {2008})}\BibitemShut {NoStop}%
\bibitem [{\citenamefont {Zârbo}\ and\ \citenamefont
  {Nikolić}(2007)}]{ZarN07}%
  \BibitemOpen
  \bibfield  {author} {\bibinfo {author} {\bibfnamefont {L.~P.}\ \bibnamefont
  {Zârbo}}\ and\ \bibinfo {author} {\bibfnamefont {B.~K.}\ \bibnamefont
  {Nikolić}},\ }\href {http://stacks.iop.org/0295-5075/80/i=4/a=47001}
  {\bibfield  {journal} {\bibinfo  {journal} {EPL (Europhysics Letters)}\
  }\textbf {\bibinfo {volume} {80}},\ \bibinfo {pages} {47001} (\bibinfo {year}
  {2007})}\BibitemShut {NoStop}%
\bibitem [{\citenamefont {McCann}(2006)}]{McC06}%
  \BibitemOpen
  \bibfield  {author} {\bibinfo {author} {\bibfnamefont {E.}~\bibnamefont
  {McCann}},\ }\href {\doibase 10.1103/PhysRevB.74.161403} {\bibfield
  {journal} {\bibinfo  {journal} {Phys. Rev. B}\ }\textbf {\bibinfo {volume}
  {74}},\ \bibinfo {pages} {161403} (\bibinfo {year} {2006})}\BibitemShut
  {NoStop}%
\bibitem [{\citenamefont {Castro}\ \emph {et~al.}(2007)\citenamefont {Castro},
  \citenamefont {Novoselov}, \citenamefont {Morozov}, \citenamefont {Peres},
  \citenamefont {dos Santos}, \citenamefont {Nilsson}, \citenamefont {Guinea},
  \citenamefont {Geim},\ and\ \citenamefont {Neto}}]{CasNN10}%
  \BibitemOpen
  \bibfield  {author} {\bibinfo {author} {\bibfnamefont {E.~V.}\ \bibnamefont
  {Castro}}, \bibinfo {author} {\bibfnamefont {K.~S.}\ \bibnamefont
  {Novoselov}}, \bibinfo {author} {\bibfnamefont {S.~V.}\ \bibnamefont
  {Morozov}}, \bibinfo {author} {\bibfnamefont {N.~M.~R.}\ \bibnamefont
  {Peres}}, \bibinfo {author} {\bibfnamefont {J.~M. B.~L.}\ \bibnamefont {dos
  Santos}}, \bibinfo {author} {\bibfnamefont {J.}~\bibnamefont {Nilsson}},
  \bibinfo {author} {\bibfnamefont {F.}~\bibnamefont {Guinea}}, \bibinfo
  {author} {\bibfnamefont {A.~K.}\ \bibnamefont {Geim}}, \ and\ \bibinfo
  {author} {\bibfnamefont {A.~H.~C.}\ \bibnamefont {Neto}},\ }\href {\doibase
  10.1103/PhysRevLett.99.216802} {\bibfield  {journal} {\bibinfo  {journal}
  {Phys. Rev. Lett.}\ }\textbf {\bibinfo {volume} {99}},\ \bibinfo {pages}
  {216802} (\bibinfo {year} {2007})}\BibitemShut {NoStop}%
\bibitem [{\citenamefont {Xu}\ \emph {et~al.}(2013)\citenamefont {Xu},
  \citenamefont {Liu}, \citenamefont {IV}, \citenamefont {Song}, \citenamefont
  {Jiang}, \citenamefont {feng Sun},\ and\ \citenamefont {Xie}}]{DonVJ13}%
  \BibitemOpen
  \bibfield  {author} {\bibinfo {author} {\bibfnamefont {D.}~\bibnamefont
  {Xu}}, \bibinfo {author} {\bibfnamefont {H.}~\bibnamefont {Liu}}, \bibinfo
  {author} {\bibfnamefont {V.~S.}\ \bibnamefont {IV}}, \bibinfo {author}
  {\bibfnamefont {J.}~\bibnamefont {Song}}, \bibinfo {author} {\bibfnamefont
  {H.}~\bibnamefont {Jiang}}, \bibinfo {author} {\bibfnamefont
  {Q.}~\bibnamefont {feng Sun}}, \ and\ \bibinfo {author} {\bibfnamefont
  {X.~C.}\ \bibnamefont {Xie}},\ }\href
  {http://stacks.iop.org/0953-8984/25/i=10/a=105303} {\bibfield  {journal}
  {\bibinfo  {journal} {Journal of Physics: Condensed Matter}\ }\textbf
  {\bibinfo {volume} {25}},\ \bibinfo {pages} {105303} (\bibinfo {year}
  {2013})}\BibitemShut {NoStop}%
\bibitem [{\citenamefont {Ohta}\ \emph {et~al.}(2006)\citenamefont {Ohta},
  \citenamefont {Bostwick}, \citenamefont {Seyller}, \citenamefont {Horn},\
  and\ \citenamefont {Rotenberg}}]{OhtBA06}%
  \BibitemOpen
  \bibfield  {author} {\bibinfo {author} {\bibfnamefont {T.}~\bibnamefont
  {Ohta}}, \bibinfo {author} {\bibfnamefont {A.}~\bibnamefont {Bostwick}},
  \bibinfo {author} {\bibfnamefont {T.}~\bibnamefont {Seyller}}, \bibinfo
  {author} {\bibfnamefont {K.}~\bibnamefont {Horn}}, \ and\ \bibinfo {author}
  {\bibfnamefont {E.}~\bibnamefont {Rotenberg}},\ }\href {\doibase
  10.1126/science.1130681} {\bibfield  {journal} {\bibinfo  {journal}
  {Science}\ }\textbf {\bibinfo {volume} {313}},\ \bibinfo {pages} {951}
  (\bibinfo {year} {2006})},\ \Eprint
  {http://arxiv.org/abs/http://www.sciencemag.org/content/313/5789/951.full.pdf}
  {http://www.sciencemag.org/content/313/5789/951.full.pdf} \BibitemShut
  {NoStop}%
\bibitem [{\citenamefont {Nilsson}\ \emph {et~al.}(2008)\citenamefont
  {Nilsson}, \citenamefont {Castro~Neto}, \citenamefont {Guinea},\ and\
  \citenamefont {Peres}}]{NilCG08}%
  \BibitemOpen
  \bibfield  {author} {\bibinfo {author} {\bibfnamefont {J.}~\bibnamefont
  {Nilsson}}, \bibinfo {author} {\bibfnamefont {A.~H.}\ \bibnamefont
  {Castro~Neto}}, \bibinfo {author} {\bibfnamefont {F.}~\bibnamefont {Guinea}},
  \ and\ \bibinfo {author} {\bibfnamefont {N.~M.~R.}\ \bibnamefont {Peres}},\
  }\href {\doibase 10.1103/PhysRevB.78.045405} {\bibfield  {journal} {\bibinfo
  {journal} {Phys. Rev. B}\ }\textbf {\bibinfo {volume} {78}},\ \bibinfo
  {pages} {045405} (\bibinfo {year} {2008})}\BibitemShut {NoStop}%
\bibitem [{\citenamefont {McCann}\ and\ \citenamefont
  {Koshino}(2013)}]{McCK13}%
  \BibitemOpen
  \bibfield  {author} {\bibinfo {author} {\bibfnamefont {E.}~\bibnamefont
  {McCann}}\ and\ \bibinfo {author} {\bibfnamefont {M.}~\bibnamefont
  {Koshino}},\ }\href {http://stacks.iop.org/0034-4885/76/i=5/a=056503}
  {\bibfield  {journal} {\bibinfo  {journal} {Reports on Progress in Physics}\
  }\textbf {\bibinfo {volume} {76}},\ \bibinfo {pages} {056503} (\bibinfo
  {year} {2013})}\BibitemShut {NoStop}%
\bibitem [{\citenamefont {Zhang}\ \emph {et~al.}(2009)\citenamefont {Zhang},
  \citenamefont {Tang}, \citenamefont {Girit}, \citenamefont {Hao},
  \citenamefont {Martin}, \citenamefont {Zettl},\ and\ \citenamefont
  {Crommie}}]{ZhaTG09}%
  \BibitemOpen
  \bibfield  {author} {\bibinfo {author} {\bibfnamefont {Y.}~\bibnamefont
  {Zhang}}, \bibinfo {author} {\bibfnamefont {T.-T.}\ \bibnamefont {Tang}},
  \bibinfo {author} {\bibfnamefont {C.}~\bibnamefont {Girit}}, \bibinfo
  {author} {\bibfnamefont {Z.}~\bibnamefont {Hao}}, \bibinfo {author}
  {\bibfnamefont {M.~C.}\ \bibnamefont {Martin}}, \bibinfo {author}
  {\bibfnamefont {A.}~\bibnamefont {Zettl}}, \ and\ \bibinfo {author}
  {\bibfnamefont {Y.~R. W.~F.}\ \bibnamefont {Crommie}, \bibfnamefont {Michael
  F. nad~Shen}},\ }\href {\doibase 10.1038/nature08105} {\bibfield  {journal}
  {\bibinfo  {journal} {Nature}\ }\textbf {\bibinfo {volume} {459}},\ \bibinfo
  {pages} {820} (\bibinfo {year} {2009})}\BibitemShut {NoStop}%
\bibitem [{\citenamefont {Abedinpour}\ \emph {et~al.}(2007)\citenamefont
  {Abedinpour}, \citenamefont {Polini}, \citenamefont {MacDonald},
  \citenamefont {Tanatar}, \citenamefont {Tosi},\ and\ \citenamefont
  {Vignale}}]{AbdPM07}%
  \BibitemOpen
  \bibfield  {author} {\bibinfo {author} {\bibfnamefont {S.~H.}\ \bibnamefont
  {Abedinpour}}, \bibinfo {author} {\bibfnamefont {M.}~\bibnamefont {Polini}},
  \bibinfo {author} {\bibfnamefont {A.~H.}\ \bibnamefont {MacDonald}}, \bibinfo
  {author} {\bibfnamefont {B.}~\bibnamefont {Tanatar}}, \bibinfo {author}
  {\bibfnamefont {M.~P.}\ \bibnamefont {Tosi}}, \ and\ \bibinfo {author}
  {\bibfnamefont {G.}~\bibnamefont {Vignale}},\ }\href {\doibase
  10.1103/PhysRevLett.99.206802} {\bibfield  {journal} {\bibinfo  {journal}
  {Phys. Rev. Lett.}\ }\textbf {\bibinfo {volume} {99}},\ \bibinfo {pages}
  {206802} (\bibinfo {year} {2007})}\BibitemShut {NoStop}%
\bibitem [{\citenamefont {Miyazaki}\ \emph {et~al.}(2010)\citenamefont
  {Miyazaki}, \citenamefont {Tsukagoshi}, \citenamefont {Kanda}, \citenamefont
  {Otani},\ and\ \citenamefont {Okada}}]{MiyTK10}%
  \BibitemOpen
  \bibfield  {author} {\bibinfo {author} {\bibfnamefont {H.}~\bibnamefont
  {Miyazaki}}, \bibinfo {author} {\bibfnamefont {K.}~\bibnamefont
  {Tsukagoshi}}, \bibinfo {author} {\bibfnamefont {A.}~\bibnamefont {Kanda}},
  \bibinfo {author} {\bibfnamefont {M.}~\bibnamefont {Otani}}, \ and\ \bibinfo
  {author} {\bibfnamefont {S.}~\bibnamefont {Okada}},\ }\href {\doibase
  10.1021/nl1015365} {\bibfield  {journal} {\bibinfo  {journal} {Nano Letters}\
  }\textbf {\bibinfo {volume} {10}},\ \bibinfo {pages} {3888} (\bibinfo {year}
  {2010})},\ \Eprint
  {http://arxiv.org/abs/http://pubs.acs.org/doi/pdf/10.1021/nl1015365}
  {http://pubs.acs.org/doi/pdf/10.1021/nl1015365} \BibitemShut {NoStop}%
\bibitem [{\citenamefont {Xia}\ \emph {et~al.}(2010)\citenamefont {Xia},
  \citenamefont {Farmer}, \citenamefont {Lin},\ and\ \citenamefont
  {Avouris}}]{XiaFL10}%
  \BibitemOpen
  \bibfield  {author} {\bibinfo {author} {\bibfnamefont {F.}~\bibnamefont
  {Xia}}, \bibinfo {author} {\bibfnamefont {D.~B.}\ \bibnamefont {Farmer}},
  \bibinfo {author} {\bibfnamefont {Y.-m.}\ \bibnamefont {Lin}}, \ and\
  \bibinfo {author} {\bibfnamefont {P.}~\bibnamefont {Avouris}},\ }\href
  {\doibase 10.1021/nl9039636} {\bibfield  {journal} {\bibinfo  {journal} {Nano
  Letters}\ }\textbf {\bibinfo {volume} {10}},\ \bibinfo {pages} {715}
  (\bibinfo {year} {2010})},\ \bibinfo {note} {pMID: 20092332},\ \Eprint
  {http://arxiv.org/abs/http://pubs.acs.org/doi/pdf/10.1021/nl9039636}
  {http://pubs.acs.org/doi/pdf/10.1021/nl9039636} \BibitemShut {NoStop}%
\bibitem [{\citenamefont {J.}\ \emph {et~al.}(2012)\citenamefont {J.},
  \citenamefont {L.}, \citenamefont {W.}, \citenamefont {Y.}, \citenamefont
  {P.}, \citenamefont {V.}, \citenamefont {M.}, \citenamefont {N.},
  \citenamefont {C.}, \citenamefont {R.}, \citenamefont {D.}, \citenamefont
  {Zhang}, \citenamefont {J.},\ and\ \citenamefont {H.}}]{Vel12}%
  \BibitemOpen
  \bibfield  {author} {\bibinfo {author} {\bibfnamefont {V.}~\bibnamefont
  {J.}}, \bibinfo {author} {\bibfnamefont {J.}~\bibnamefont {L.}}, \bibinfo
  {author} {\bibfnamefont {B.}~\bibnamefont {W.}}, \bibinfo {author}
  {\bibfnamefont {L.}~\bibnamefont {Y.}}, \bibinfo {author} {\bibfnamefont
  {K.}~\bibnamefont {P.}}, \bibinfo {author} {\bibfnamefont {A.}~\bibnamefont
  {V.}}, \bibinfo {author} {\bibfnamefont {B.}~\bibnamefont {M.}}, \bibinfo
  {author} {\bibfnamefont {L.~C.}\ \bibnamefont {N.}}, \bibinfo {author}
  {\bibfnamefont {V.}~\bibnamefont {C.}}, \bibinfo {author} {\bibfnamefont
  {S.}~\bibnamefont {R.}}, \bibinfo {author} {\bibfnamefont {S.}~\bibnamefont
  {D.}}, \bibinfo {author} {\bibfnamefont {F.}~\bibnamefont {Zhang}}, \bibinfo
  {author} {\bibfnamefont {J.}~\bibnamefont {J.}}, \ and\ \bibinfo {author}
  {\bibfnamefont {M.~A.}\ \bibnamefont {H.}},\ }\href {\doibase
  dx.doi.org/10.1038/nnano.2011.251} {\bibfield  {journal} {\bibinfo  {journal}
  {Nat Nano}\ }\textbf {\bibinfo {volume} {7}},\ \bibinfo {pages} {156}
  (\bibinfo {year} {2012})}\BibitemShut {NoStop}%
\bibitem [{\citenamefont {Qiao}\ \emph {et~al.}(2011)\citenamefont {Qiao},
  \citenamefont {Jung}, \citenamefont {Niu},\ and\ \citenamefont
  {MacDonald}}]{QiaJN11}%
  \BibitemOpen
  \bibfield  {author} {\bibinfo {author} {\bibfnamefont {Z.}~\bibnamefont
  {Qiao}}, \bibinfo {author} {\bibfnamefont {J.}~\bibnamefont {Jung}}, \bibinfo
  {author} {\bibfnamefont {Q.}~\bibnamefont {Niu}}, \ and\ \bibinfo {author}
  {\bibfnamefont {A.~H.}\ \bibnamefont {MacDonald}},\ }\href {\doibase
  10.1021/nl201941f} {\bibfield  {journal} {\bibinfo  {journal} {Nano Letters}\
  }\textbf {\bibinfo {volume} {11}},\ \bibinfo {pages} {3453} (\bibinfo {year}
  {2011})},\ \Eprint
  {http://arxiv.org/abs/http://pubs.acs.org/doi/pdf/10.1021/nl201941f}
  {http://pubs.acs.org/doi/pdf/10.1021/nl201941f} \BibitemShut {NoStop}%
\bibitem [{\citenamefont {Li}\ \emph {et~al.}(2010)\citenamefont {Li},
  \citenamefont {Zhang},\ and\ \citenamefont {Xiao}}]{LiZX10}%
  \BibitemOpen
  \bibfield  {author} {\bibinfo {author} {\bibfnamefont {X.}~\bibnamefont
  {Li}}, \bibinfo {author} {\bibfnamefont {Z.}~\bibnamefont {Zhang}}, \ and\
  \bibinfo {author} {\bibfnamefont {D.}~\bibnamefont {Xiao}},\ }\href {\doibase
  10.1103/PhysRevB.81.195402} {\bibfield  {journal} {\bibinfo  {journal} {Phys.
  Rev. B}\ }\textbf {\bibinfo {volume} {81}},\ \bibinfo {pages} {195402}
  (\bibinfo {year} {2010})}\BibitemShut {NoStop}%
\bibitem [{\citenamefont {Kim}\ \emph {et~al.}(2013)\citenamefont {Kim},
  \citenamefont {Kim}, \citenamefont {Walter}, \citenamefont {Seyller},
  \citenamefont {Yeom}, \citenamefont {Rotenberg},\ and\ \citenamefont
  {Bostwick}}]{KimKW13}%
  \BibitemOpen
  \bibfield  {author} {\bibinfo {author} {\bibfnamefont {K.~S.}\ \bibnamefont
  {Kim}}, \bibinfo {author} {\bibfnamefont {T.-H.}\ \bibnamefont {Kim}},
  \bibinfo {author} {\bibfnamefont {A.~L.}\ \bibnamefont {Walter}}, \bibinfo
  {author} {\bibfnamefont {T.}~\bibnamefont {Seyller}}, \bibinfo {author}
  {\bibfnamefont {H.~W.}\ \bibnamefont {Yeom}}, \bibinfo {author}
  {\bibfnamefont {E.}~\bibnamefont {Rotenberg}}, \ and\ \bibinfo {author}
  {\bibfnamefont {A.}~\bibnamefont {Bostwick}},\ }\href {\doibase
  10.1103/PhysRevLett.110.036804} {\bibfield  {journal} {\bibinfo  {journal}
  {Phys. Rev. Lett.}\ }\textbf {\bibinfo {volume} {110}},\ \bibinfo {pages}
  {036804} (\bibinfo {year} {2013})}\BibitemShut {NoStop}%
\bibitem [{\citenamefont {Nakada}\ \emph {et~al.}(1996)\citenamefont {Nakada},
  \citenamefont {Fujita}, \citenamefont {Dresselhaus},\ and\ \citenamefont
  {Dresselhaus}}]{NakFD96}%
  \BibitemOpen
  \bibfield  {author} {\bibinfo {author} {\bibfnamefont {K.}~\bibnamefont
  {Nakada}}, \bibinfo {author} {\bibfnamefont {M.}~\bibnamefont {Fujita}},
  \bibinfo {author} {\bibfnamefont {G.}~\bibnamefont {Dresselhaus}}, \ and\
  \bibinfo {author} {\bibfnamefont {M.~S.}\ \bibnamefont {Dresselhaus}},\
  }\href {\doibase 10.1103/PhysRevB.54.17954} {\bibfield  {journal} {\bibinfo
  {journal} {Phys. Rev. B}\ }\textbf {\bibinfo {volume} {54}},\ \bibinfo
  {pages} {17954} (\bibinfo {year} {1996})}\BibitemShut {NoStop}%
\bibitem [{\citenamefont {Lima}\ \emph {et~al.}(2010)\citenamefont {Lima},
  \citenamefont {da~Silva},\ and\ \citenamefont {Fazzio}}]{LimSF10}%
  \BibitemOpen
  \bibfield  {author} {\bibinfo {author} {\bibfnamefont {M.~P.}\ \bibnamefont
  {Lima}}, \bibinfo {author} {\bibfnamefont {A.~J.~R.}\ \bibnamefont
  {da~Silva}}, \ and\ \bibinfo {author} {\bibfnamefont {A.}~\bibnamefont
  {Fazzio}},\ }\href {\doibase 10.1103/PhysRevB.81.045430} {\bibfield
  {journal} {\bibinfo  {journal} {Phys. Rev. B}\ }\textbf {\bibinfo {volume}
  {81}},\ \bibinfo {pages} {045430} (\bibinfo {year} {2010})}\BibitemShut
  {NoStop}%
\bibitem [{\citenamefont {Lv}\ and\ \citenamefont {Li}(2012)}]{LvL12}%
  \BibitemOpen
  \bibfield  {author} {\bibinfo {author} {\bibfnamefont {S.-H.}\ \bibnamefont
  {Lv}}\ and\ \bibinfo {author} {\bibfnamefont {Y.-X.}\ \bibnamefont {Li}},\
  }\href {\doibase http://dx.doi.org/10.1063/1.4747927} {\bibfield  {journal}
  {\bibinfo  {journal} {Journal of Applied Physics}\ }\textbf {\bibinfo
  {volume} {112}},\ \bibinfo {eid} {053701} (\bibinfo {year}
  {2012})}\BibitemShut {NoStop}%
\bibitem [{\citenamefont {Wang}\ \emph {et~al.}(2011)\citenamefont {Wang},
  \citenamefont {Song}, \citenamefont {Lee}, \citenamefont {Tang},
  \citenamefont {Lang}, \citenamefont {Zeng}, \citenamefont {Xu}, \citenamefont
  {Zhou},\ and\ \citenamefont {Wang}}]{WanSL11}%
  \BibitemOpen
  \bibfield  {author} {\bibinfo {author} {\bibfnamefont {M.}~\bibnamefont
  {Wang}}, \bibinfo {author} {\bibfnamefont {E.~B.}\ \bibnamefont {Song}},
  \bibinfo {author} {\bibfnamefont {S.}~\bibnamefont {Lee}}, \bibinfo {author}
  {\bibfnamefont {J.}~\bibnamefont {Tang}}, \bibinfo {author} {\bibfnamefont
  {M.}~\bibnamefont {Lang}}, \bibinfo {author} {\bibfnamefont {C.}~\bibnamefont
  {Zeng}}, \bibinfo {author} {\bibfnamefont {G.}~\bibnamefont {Xu}}, \bibinfo
  {author} {\bibfnamefont {Y.}~\bibnamefont {Zhou}}, \ and\ \bibinfo {author}
  {\bibfnamefont {K.~L.}\ \bibnamefont {Wang}},\ }\href {\doibase
  10.1021/nn2027566} {\bibfield  {journal} {\bibinfo  {journal} {ACS Nano}\
  }\textbf {\bibinfo {volume} {5}},\ \bibinfo {pages} {8769} (\bibinfo {year}
  {2011})},\ \Eprint
  {http://arxiv.org/abs/http://pubs.acs.org/doi/pdf/10.1021/nn2027566}
  {http://pubs.acs.org/doi/pdf/10.1021/nn2027566} \BibitemShut {NoStop}%
\bibitem [{\citenamefont {Kuzmenko}\ \emph {et~al.}(2009)\citenamefont
  {Kuzmenko}, \citenamefont {Crassee}, \citenamefont {van~der Marel},
  \citenamefont {Blake},\ and\ \citenamefont {Novoselov}}]{KuzCV09}%
  \BibitemOpen
  \bibfield  {author} {\bibinfo {author} {\bibfnamefont {A.~B.}\ \bibnamefont
  {Kuzmenko}}, \bibinfo {author} {\bibfnamefont {I.}~\bibnamefont {Crassee}},
  \bibinfo {author} {\bibfnamefont {D.}~\bibnamefont {van~der Marel}}, \bibinfo
  {author} {\bibfnamefont {P.}~\bibnamefont {Blake}}, \ and\ \bibinfo {author}
  {\bibfnamefont {K.~S.}\ \bibnamefont {Novoselov}},\ }\href {\doibase
  10.1103/PhysRevB.80.165406} {\bibfield  {journal} {\bibinfo  {journal} {Phys.
  Rev. B}\ }\textbf {\bibinfo {volume} {80}},\ \bibinfo {pages} {165406}
  (\bibinfo {year} {2009})}\BibitemShut {NoStop}%
\bibitem [{\citenamefont {Datta}(1999)}]{Datta99}%
  \BibitemOpen
  \bibfield  {author} {\bibinfo {author} {\bibfnamefont {S.}~\bibnamefont
  {Datta}},\ }\href@noop {} {\emph {\bibinfo {title} {Electronic Transport in
  Mesoscopic Systems}}}\ (\bibinfo  {publisher} {Cambridge University Press},\
  \bibinfo {address} {Cambridge},\ \bibinfo {year} {1999})\BibitemShut
  {NoStop}%
\bibitem [{\citenamefont {Ferry}\ and\ \citenamefont {Goodnick}(1999)}]{FG99}%
  \BibitemOpen
  \bibfield  {author} {\bibinfo {author} {\bibfnamefont {D.~K.}\ \bibnamefont
  {Ferry}}\ and\ \bibinfo {author} {\bibfnamefont {S.~M.}\ \bibnamefont
  {Goodnick}},\ }\href@noop {} {\emph {\bibinfo {title} {Transport in
  Nanostructures}}},\ \bibinfo {series} {Cambridge Studies in Semiconductor
  Physics \& Microelectronic Engineering}, Vol.~\bibinfo {volume} {6}\
  (\bibinfo  {publisher} {Cambridge University Press},\ \bibinfo {year}
  {1999})\BibitemShut {NoStop}%
\bibitem [{\citenamefont {Lewenkopf}\ and\ \citenamefont
  {Mucciolo}(2013)}]{LewM13}%
  \BibitemOpen
  \bibfield  {author} {\bibinfo {author} {\bibfnamefont {C.~H.}\ \bibnamefont
  {Lewenkopf}}\ and\ \bibinfo {author} {\bibfnamefont {E.~R.}\ \bibnamefont
  {Mucciolo}},\ }\href {\doibase 10.1007/s10825-013-0458-7} {\bibfield
  {journal} {\bibinfo  {journal} {J. Comput. Electron.}\ }\textbf {\bibinfo
  {volume} {12}},\ \bibinfo {pages} {203} (\bibinfo {year} {2013})}\BibitemShut
  {NoStop}%
\bibitem [{\citenamefont {Bahamon}\ \emph {et~al.}(2011)\citenamefont
  {Bahamon}, \citenamefont {Pereira},\ and\ \citenamefont {Schulz}}]{BahPS11}%
  \BibitemOpen
  \bibfield  {author} {\bibinfo {author} {\bibfnamefont {D.~A.}\ \bibnamefont
  {Bahamon}}, \bibinfo {author} {\bibfnamefont {A.~L.~C.}\ \bibnamefont
  {Pereira}}, \ and\ \bibinfo {author} {\bibfnamefont {P.~A.}\ \bibnamefont
  {Schulz}},\ }\href {\doibase 10.1103/PhysRevB.83.155436} {\bibfield
  {journal} {\bibinfo  {journal} {Phys. Rev. B}\ }\textbf {\bibinfo {volume}
  {83}},\ \bibinfo {pages} {155436} (\bibinfo {year} {2011})}\BibitemShut
  {NoStop}%
\bibitem [{\citenamefont {Haug}(2008)}]{HauJ08}%
  \BibitemOpen
  \bibfield  {author} {\bibinfo {author} {\bibfnamefont {J.~A.-P.}\
  \bibnamefont {Haug}, \bibfnamefont {H.}},\ }\href@noop {} {\emph {\bibinfo
  {title} {Quantum Kinetics in Transport and Optics of Semiconductors}}}\
  (\bibinfo  {publisher} {Springer Berlin Heidelberg},\ \bibinfo {year}
  {2008})\BibitemShut {NoStop}%
\bibitem [{\citenamefont {Cresti}\ \emph {et~al.}(2003)\citenamefont {Cresti},
  \citenamefont {Farchioni}, \citenamefont {Grosso},\ and\ \citenamefont
  {Parravicini}}]{CreFG03}%
  \BibitemOpen
  \bibfield  {author} {\bibinfo {author} {\bibfnamefont {A.}~\bibnamefont
  {Cresti}}, \bibinfo {author} {\bibfnamefont {R.}~\bibnamefont {Farchioni}},
  \bibinfo {author} {\bibfnamefont {G.}~\bibnamefont {Grosso}}, \ and\ \bibinfo
  {author} {\bibfnamefont {G.~P.}\ \bibnamefont {Parravicini}},\ }\href
  {\doibase 10.1103/PhysRevB.68.075306} {\bibfield  {journal} {\bibinfo
  {journal} {Phys. Rev. B}\ }\textbf {\bibinfo {volume} {68}},\ \bibinfo
  {pages} {075306} (\bibinfo {year} {2003})}\BibitemShut {NoStop}%
\bibitem [{\citenamefont {Metalidis}\ and\ \citenamefont
  {Bruno}(2005)}]{MetB05}%
  \BibitemOpen
  \bibfield  {author} {\bibinfo {author} {\bibfnamefont {G.}~\bibnamefont
  {Metalidis}}\ and\ \bibinfo {author} {\bibfnamefont {P.}~\bibnamefont
  {Bruno}},\ }\href {\doibase 10.1103/PhysRevB.72.235304} {\bibfield  {journal}
  {\bibinfo  {journal} {Phys. Rev. B}\ }\textbf {\bibinfo {volume} {72}},\
  \bibinfo {pages} {235304} (\bibinfo {year} {2005})}\BibitemShut {NoStop}%
\bibitem [{\citenamefont {Castro}\ \emph {et~al.}(2010)\citenamefont {Castro},
  \citenamefont {Novoselov}, \citenamefont {Morozov}, \citenamefont {Peres},
  \citenamefont {dos Santos}, \citenamefont {Nilsson}, \citenamefont {Guinea},
  \citenamefont {Geim},\ and\ \citenamefont {Neto}}]{CasNM10}%
  \BibitemOpen
  \bibfield  {author} {\bibinfo {author} {\bibfnamefont {E.~V.}\ \bibnamefont
  {Castro}}, \bibinfo {author} {\bibfnamefont {K.~S.}\ \bibnamefont
  {Novoselov}}, \bibinfo {author} {\bibfnamefont {S.~V.}\ \bibnamefont
  {Morozov}}, \bibinfo {author} {\bibfnamefont {N.~M.~R.}\ \bibnamefont
  {Peres}}, \bibinfo {author} {\bibfnamefont {J.~M. B.~L.}\ \bibnamefont {dos
  Santos}}, \bibinfo {author} {\bibfnamefont {J.}~\bibnamefont {Nilsson}},
  \bibinfo {author} {\bibfnamefont {F.}~\bibnamefont {Guinea}}, \bibinfo
  {author} {\bibfnamefont {A.~K.}\ \bibnamefont {Geim}}, \ and\ \bibinfo
  {author} {\bibfnamefont {A.~H.~C.}\ \bibnamefont {Neto}},\ }\href
  {http://stacks.iop.org/0953-8984/22/i=17/a=175503} {\bibfield  {journal}
  {\bibinfo  {journal} {Journal of Physics: Condensed Matter}\ }\textbf
  {\bibinfo {volume} {22}},\ \bibinfo {pages} {175503} (\bibinfo {year}
  {2010})}\BibitemShut {NoStop}%
\bibitem [{\citenamefont {Yao}\ \emph {et~al.}(2009)\citenamefont {Yao},
  \citenamefont {Yang},\ and\ \citenamefont {Niu}}]{YaoYN09}%
  \BibitemOpen
  \bibfield  {author} {\bibinfo {author} {\bibfnamefont {W.}~\bibnamefont
  {Yao}}, \bibinfo {author} {\bibfnamefont {S.~A.}\ \bibnamefont {Yang}}, \
  and\ \bibinfo {author} {\bibfnamefont {Q.}~\bibnamefont {Niu}},\ }\href
  {\doibase 10.1103/PhysRevLett.102.096801} {\bibfield  {journal} {\bibinfo
  {journal} {Phys. Rev. Lett.}\ }\textbf {\bibinfo {volume} {102}},\ \bibinfo
  {pages} {096801} (\bibinfo {year} {2009})}\BibitemShut {NoStop}%
\bibitem [{\citenamefont {Rhim}\ and\ \citenamefont {Moon}(2008)}]{RhiK08}%
  \BibitemOpen
  \bibfield  {author} {\bibinfo {author} {\bibfnamefont {J.-W.}\ \bibnamefont
  {Rhim}}\ and\ \bibinfo {author} {\bibfnamefont {K.}~\bibnamefont {Moon}},\
  }\href {http://stacks.iop.org/0953-8984/20/i=36/a=365202} {\bibfield
  {journal} {\bibinfo  {journal} {Journal of Physics: Condensed Matter}\
  }\textbf {\bibinfo {volume} {20}},\ \bibinfo {pages} {365202} (\bibinfo
  {year} {2008})}\BibitemShut {NoStop}%
\bibitem [{\citenamefont {Koshino}\ and\ \citenamefont {Ando}(2006)}]{KosA06}%
  \BibitemOpen
  \bibfield  {author} {\bibinfo {author} {\bibfnamefont {M.}~\bibnamefont
  {Koshino}}\ and\ \bibinfo {author} {\bibfnamefont {T.}~\bibnamefont {Ando}},\
  }\href {\doibase 10.1103/PhysRevB.73.245403} {\bibfield  {journal} {\bibinfo
  {journal} {Phys. Rev. B}\ }\textbf {\bibinfo {volume} {73}},\ \bibinfo
  {pages} {245403} (\bibinfo {year} {2006})}\BibitemShut {NoStop}%
\bibitem [{\citenamefont {Ramasubramaniam}\ \emph {et~al.}(2011)\citenamefont
  {Ramasubramaniam}, \citenamefont {Naveh},\ and\ \citenamefont
  {Towe}}]{RamNT11}%
  \BibitemOpen
  \bibfield  {author} {\bibinfo {author} {\bibfnamefont {A.}~\bibnamefont
  {Ramasubramaniam}}, \bibinfo {author} {\bibfnamefont {D.}~\bibnamefont
  {Naveh}}, \ and\ \bibinfo {author} {\bibfnamefont {E.}~\bibnamefont {Towe}},\
  }\href {\doibase 10.1021/nl1039499} {\bibfield  {journal} {\bibinfo
  {journal} {Nano Letters}\ }\textbf {\bibinfo {volume} {11}},\ \bibinfo
  {pages} {1070} (\bibinfo {year} {2011})},\ \Eprint
  {http://arxiv.org/abs/http://pubs.acs.org/doi/pdf/10.1021/nl1039499}
  {http://pubs.acs.org/doi/pdf/10.1021/nl1039499} \BibitemShut {NoStop}%
\bibitem [{\citenamefont {Mucciolo}\ \emph {et~al.}(2009)\citenamefont
  {Mucciolo}, \citenamefont {Castro~Neto},\ and\ \citenamefont
  {Lewenkopf}}]{MucCL09}%
  \BibitemOpen
  \bibfield  {author} {\bibinfo {author} {\bibfnamefont {E.~R.}\ \bibnamefont
  {Mucciolo}}, \bibinfo {author} {\bibfnamefont {A.~H.}\ \bibnamefont
  {Castro~Neto}}, \ and\ \bibinfo {author} {\bibfnamefont {C.~H.}\ \bibnamefont
  {Lewenkopf}},\ }\href {\doibase 10.1103/PhysRevB.79.075407} {\bibfield
  {journal} {\bibinfo  {journal} {Phys. Rev. B}\ }\textbf {\bibinfo {volume}
  {79}},\ \bibinfo {pages} {075407} (\bibinfo {year} {2009})}\BibitemShut
  {NoStop}%
\bibitem [{\citenamefont {Cresti}\ and\ \citenamefont {Roche}(2009)}]{CreR09}%
  \BibitemOpen
  \bibfield  {author} {\bibinfo {author} {\bibfnamefont {A.}~\bibnamefont
  {Cresti}}\ and\ \bibinfo {author} {\bibfnamefont {S.}~\bibnamefont {Roche}},\
  }\href {http://stacks.iop.org/1367-2630/11/i=9/a=095004} {\bibfield
  {journal} {\bibinfo  {journal} {New Journal of Physics}\ }\textbf {\bibinfo
  {volume} {11}},\ \bibinfo {pages} {095004} (\bibinfo {year}
  {2009})}\BibitemShut {NoStop}%
\bibitem [{\citenamefont {Mucciolo}\ and\ \citenamefont
  {Lewenkopf}(2010)}]{MucL10}%
  \BibitemOpen
  \bibfield  {author} {\bibinfo {author} {\bibfnamefont {E.~R.}\ \bibnamefont
  {Mucciolo}}\ and\ \bibinfo {author} {\bibfnamefont {C.~H.}\ \bibnamefont
  {Lewenkopf}},\ }\href {http://stacks.iop.org/0953-8984/22/i=27/a=273201}
  {\bibfield  {journal} {\bibinfo  {journal} {Journal of Physics: Condensed
  Matter}\ }\textbf {\bibinfo {volume} {22}},\ \bibinfo {pages} {273201}
  (\bibinfo {year} {2010})}\BibitemShut {NoStop}%
\bibitem [{\citenamefont {Pereira}\ and\ \citenamefont
  {Schulz}(2008)}]{PerS08b}%
  \BibitemOpen
  \bibfield  {author} {\bibinfo {author} {\bibfnamefont {A.~L.~C.}\
  \bibnamefont {Pereira}}\ and\ \bibinfo {author} {\bibfnamefont {P.~A.}\
  \bibnamefont {Schulz}},\ }\href {\doibase 10.1103/PhysRevB.77.075416}
  {\bibfield  {journal} {\bibinfo  {journal} {Phys. Rev. B}\ }\textbf {\bibinfo
  {volume} {77}},\ \bibinfo {pages} {075416} (\bibinfo {year}
  {2008})}\BibitemShut {NoStop}%
\bibitem [{\citenamefont {Sasaki}\ \emph {et~al.}(2006)\citenamefont {Sasaki},
  \citenamefont {Murakami},\ and\ \citenamefont {Saito}}]{SasMS06}%
  \BibitemOpen
  \bibfield  {author} {\bibinfo {author} {\bibfnamefont {K.}~\bibnamefont
  {Sasaki}}, \bibinfo {author} {\bibfnamefont {S.}~\bibnamefont {Murakami}}, \
  and\ \bibinfo {author} {\bibfnamefont {R.}~\bibnamefont {Saito}},\ }\href
  {\doibase http://dx.doi.org/10.1063/1.2181274} {\bibfield  {journal}
  {\bibinfo  {journal} {Applied Physics Letters}\ }\textbf {\bibinfo {volume}
  {88}},\ \bibinfo {eid} {113110} (\bibinfo {year} {2006})}\BibitemShut
  {NoStop}%
\end{thebibliography}
%\nolinenumbers
\end{document}